\documentclass{bmcart}

\usepackage{amsthm,amsmath}
\usepackage{graphicx}
\usepackage{multirow}
\usepackage{url}
\usepackage[utf8]{inputenc} 

\startlocaldefs
\endlocaldefs

\begin{document}

\begin{frontmatter}

\begin{fmbox}
\dochead{Research}

\title{Modeling International Mobility using Roaming Cell Phone Traces during COVID-19 Pandemic}

\author[
   addressref={aff1,aff2},                   
   corref={aff1},                       
   noteref={n1},                        
   email={mluca@fbk.eu}   
]{\inits{ML}\fnm{Massimiliano} \snm{Luca}}
\author[
   addressref={aff1},
   email={lepri@fbk.eu}
]{\inits{BL}\fnm{Bruno} \snm{Lepri}}
\author[
   addressref={aff4},
   email={}
]{\inits{EM}\fnm{Enrique} \snm{Frias-Martinez}}
\author[
   addressref={aff3},
   email={andra.lutu@telefonica.com}
]{\inits{AL}\fnm{Andra} \snm{Lutu}}

\address[id=aff1]{
  \orgname{Bruno Kessler Foundation}, 
  \city{Trento},                              
  \cny{Italy}                                    
}
\address[id=aff2]{%
  \orgname{Free University of Bolzano},
  \city{Bolzano},
  \cny{Italy}
}

\address[id=aff3]{%
  \orgname{Telefonica Research},
  \city{Madrid},
  \cny{Spain}
}
 \address[id=aff4]{%
  \orgname{Universidad Camilo Jose Cela, CAILAB},
  \city{Madrid},
  \cny{Spain}
}

\begin{artnotes}
\note[id=n1]{Work done while at Telefónica Research} 
\end{artnotes}

\end{fmbox}

\begin{abstractbox}

\begin{abstract} 
Most of the studies related to human mobility are focused on intra-country mobility. However, there are many scenarios (e.g., spreading diseases, migration) in which timely data on international commuters are vital. Mobile phones represent a unique opportunity to monitor international mobility flows in a timely manner and with proper spatial aggregation. This work proposes using roaming data generated by mobile phones to model incoming and outgoing international mobility. We use the gravity and radiation models to capture mobility flows before and during the introduction of non-pharmaceutical interventions. However, traditional models have some limitations: for instance, mobility restrictions are not explicitly captured and may play a crucial role. To overtake such limitations, we propose the COVID Gravity Model (CGM), namely an extension of the traditional gravity model that is tailored for the pandemic scenario. This proposed approach overtakes, in terms of accuracy, the traditional models by 126.9\% for incoming mobility and by 63.9\% when modeling outgoing mobility flows. 

\end{abstract}


\end{abstractbox}
\end{frontmatter}

\section{Introduction}
In modern societies, understanding international human mobility is crucial under multiple perspectives \cite{khanna2021}. For instance, international mobility is strictly related to many of the United Nations' sustainable development goals (SDGs), such as the reduction of global inequalities, the design and development of sustainable communities, the worldwide diffusion of innovation, and others \cite{khanna2021,un_sdg_book}.

The rapid diffusion of technologies such as mobile phones, devices with GPS receivers, social media (i.e., geo-tagged posts) generate an enormous amount of data that we can utilize to investigate human movements \cite{gonzalez2008,schneider2013,blondel2015,comito2016,liao2019,luca2021survey,barbosa2018human,alessandretti2020,schlapfer2021}.

\begin{figure}[!t]
	\centering
    \includegraphics[width=0.9\linewidth]{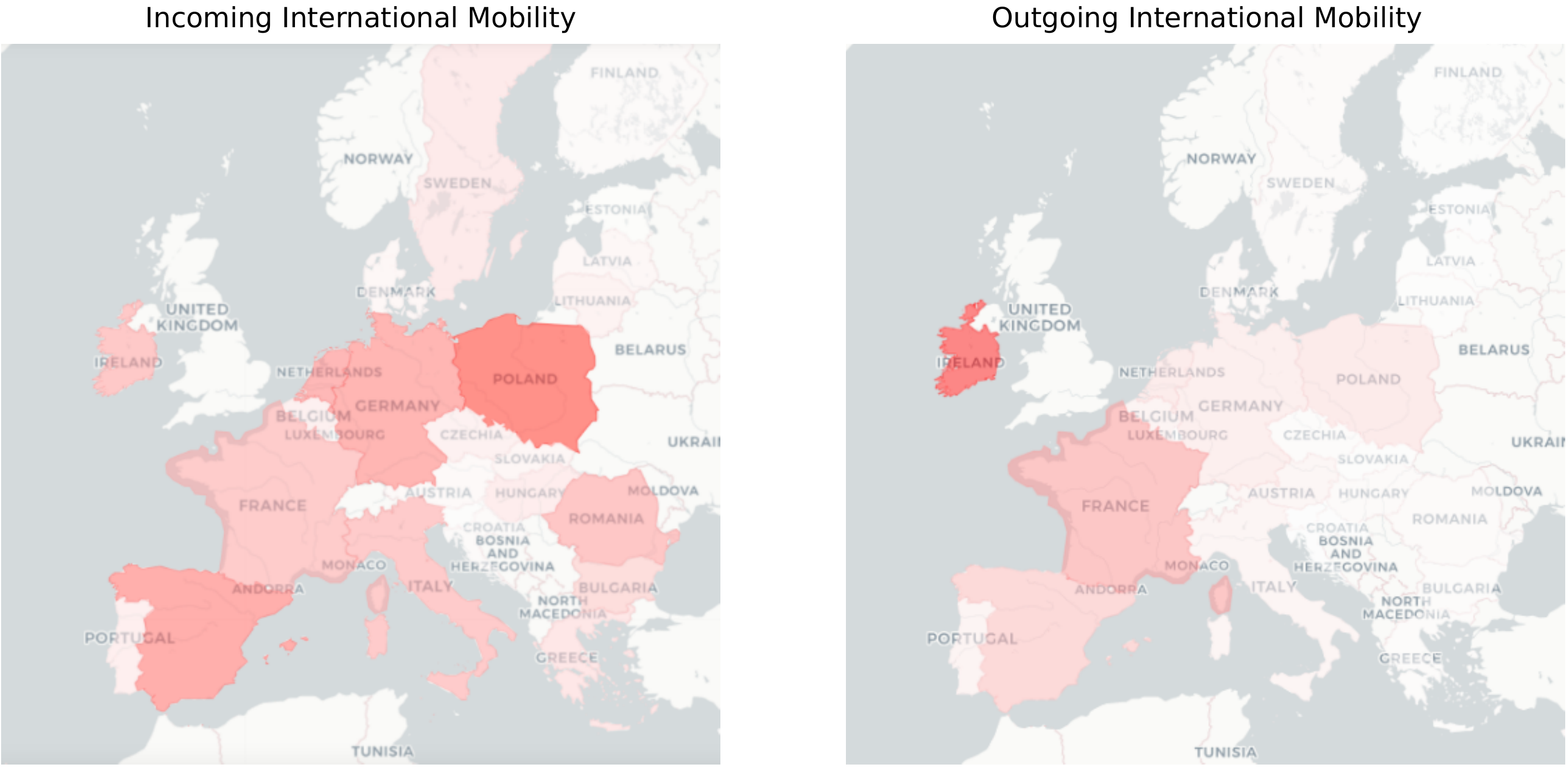}
\caption{An example of international mobility flows going from EU countries to UK (left) and from UK to EU (right). The plots refer to the $5^{th}$ of March, a business day before the introduction of mobility restrictions and other non-pharmaceutical interventions in European countries. As lighter is the red as lower is the flow. On this specific day, commuters from Poland are the ones more actively traveling to UK.}
	\label{fig:map_flows}
\end{figure}

While human mobility has been widely investigated at country and city scales, there are fewer studies regarding mobility across national borders.
In such cases, official statistics (e.g., air passenger data) have been widely used both to understand mobility trends and types of mobility of international travelers \cite{gabrielli2019dissecting,shepherd2021domestic,soria2021leveraging}, and in the context of the COVID-19 pandemic to investigate the effectiveness of non-pharmaceutical interventions (NPIs) such as international travel restrictions, to model the spreading of the disease, to measure the social and economic impact of COVID-19, and to analyze the spreading of new variants \cite{lai2021preliminary,iacus2020estimating, wolle2021stochastic,chinazzi2020effect,iacus2020flight,lemey2021untangling,kubota2020multiple,lucchini2021living}. 
Also social-media data have been used as an alternative data source to estimate international migration  \cite{alexander2020combining,rampazzo2021,zagheni2014inferring,zagheni2017leveraging,spyratos2019quantifying}. 
However, the previously mentioned works rely on data sources with some intrinsic limitations.
Social media suffer from self-selection biases. 
For example, some social media may be widely used by people of a certain age while not capturing other age bins. 
Official statistics are reliable and precise but report a limited amount of international commuters, usually those traveling with a specific mode of transport (e.g., air passengers). Moreover, statistics are generally published with significant delays. When dealing with some social issues, such as migration and disease diffusion, working with data sources that are not timely reported represents a considerable limitation. 

Hence, using mobile phone data to quantify international commuters may represent a potential solution to the challenges mentioned above.
Mobile phone data have been rarely used to deal with international commuters \cite{altin2021megastar,ahas2007mobile,nilbe2014evaluating,luca2021leveraging}. Also, in such cases the analyses were more related to the presence of mobile phones with SIMs registered in other countries more than to the commuting behaviour itself. There are also some recent works that use roaming data to predict imported COVID-19 cases \cite{choi2020forecasting,kim2020hi}. 

In this study, we use pseudo-anonymized and aggregated mobile phone data collected from a large mobile operator in UK (with a 28\% market share in 2020) to model incoming and outgoing human mobility before and during the pandemic. More precisely, incoming mobility corresponds to the number of new foreign mobile phone SIMs (i.e., SIMs that were not connected to the network of the operator the day before) while outgoing mobility is the number of mobile phone SIMs registered in the UK that travel abroad.

It has been shown that the gravity and the radiation models can efficiently model mobility in normal times \cite{barbosa2018human}. However it is not clear to what extent such models can describe international human mobility during COVID-19 pandemic. In this sense, we highlight some limitations of the gravity model and we propose an extended version named COVID Gravity Model (CGM). In CGM, we modify the deterrence function in order to take into consideration also the mobility restrictions imposed by the governments.

In summary, our contributions are as follows: 
\begin{itemize}
    \item We present the use of roaming data to model international mobility after assessing their validity by measuring the synchronicity with air traffic statistics { using well-known techniques based on Pearson correlation}.
    \item We evaluate the performance of gravity and radiation models to capture international mobility prior March 2020 and under COVID-19 restrictions.
    \item We highlight some limitations of the traditional gravity model in modeling mobility during the pandemic, and we propose COVID Gravity Model to take into account the restrictions of the analyzed countries in order to better capturing international human mobility during the COVID-19 pandemic. 
    
\end{itemize}

More specifically, in Section \ref{sec:dataset}, we first describe the dataset with a particular focus on the roaming activities (e.g., the activity of a foreign mobile phone connected to the local network or of a company SIM card connected to a foreign operator). Then, we show the process followed to extract international mobility patterns from mobile phone data.
 
In Section \ref{sec:modeling}, we show how international mobility can be modeled using a gravity model (Section \ref{sec:gravity}) and a radiation model (Section \ref{sec:radiation}). We then highlight some limitations of the traditional gravity model and in Section \ref{sec:adaptation} we propose the COVID Gravity Model (CGM) as a potential solution. 

In Section \ref{sec:results}, first, we evaluate the synchronicity between mobile phone data and air traffic data to assess the validity of roaming data (Section \ref{sec:assessing}). Next, we show the performances of the gravity and radiation models on roaming data (see Section \ref{sec:results_evaluation_gr}), while in Section \ref{sec:cgm_results} we report the performances of the COVID Gravity Model.

Finally, in Section \ref{sec:discussion} we discuss the implications and limitations of our study, and in Section \ref{sec:conclusions} we draw some conclusions and propose some future directions.

\section{Materials and Methods}
In this Section, we first present the dataset used in this study and how it was collected (Section \ref{sec:dataset}). We then introduce the gravity model and the radiation model as ways to capture human mobility (Section \ref{sec:modeling}). We also discuss the methodology behind the COVID Gravity Model (CGM) and why an extended gravity model is needed to better capture international mobility during the COVID-19 pandemic (Section \ref{sec:adaptation}). Finally, we briefly discuss the evaluation metrics adopted to evaluate the models (Section \ref{sec:evaluation_metric}).

\subsection{Dataset}
\label{sec:dataset}

Here, we describe the measurement infrastructure we leverage to collect network data from one of the largest commercial mobile network operators (MNOs) in UK, with 27.2 million subscribers as of May 2021. 
In particular, we detail the dataset we have built and the metrics we use to capture the international activity of smartphone devices. 

{
\subsubsection{Measurement Infrastructure}
In this study, we use a passive measurement approach to retrieve some anonymized information about the devices attached to the antennas of the mobile network operator that provided the data.
Each measurement carries the (1) anonymized user ID, (2) the SIM mobile country code (MCC) and mobile network code (MNC), (3) the first eight digits of the device International Mobile Equipment Identity (IMEI), (4) the timestamp, and other information. 
We also collect a device's unique ID assigned by the Global System for Mobile Communications Association that describes some properties of the device like manufacturer, brand and model name, operating system, radio bands supported, etc. In this way,  we can distinguish between smartphones (likely used as primary devices by mobile users) and Internet of Things devices. In this study, we use only the measurements related to smartphones. Additional information on the measurement infrastructure can be found in \cite{lutu2020characterization}.
}

\subsubsection{International Patterns Extraction}
Mobile phones are an ubiquitous technology that has been rapidly adopted worldwide \cite{itu2019}. 
Most of the people traveling within the same nation and internationally bring with them \textit{at least} a device that uses Radio Base Stations (RBSs) to interact with other devices (e.g., send/receive calls/messages and connect to the internet). 
Whenever people traveling with connected devices cross a border, their devices need to connect to the radio network of another (local) operator to continue working correctly. For example, a person with a mobile phone traveling from Italy to UK will have to connect to a UK telecommunication operator network. The telecommunication operator will collect information about that device, including the country where the connected SIM is registered. The latter can be extracted using the MCC, a three-digit code that allows us to identify the origin of the SIM \cite{itu2019}.
While using the generated data we can quantify the incoming international mobility, it is also possible to capture outgoing international mobility as telecommunication operators are aware of their SIMs connected to other operators' networks.

In this study, to quantify international mobility, we are interested in counting (1) the number of foreign mobile phones connected to operators' network per day as a proxy of incoming international mobility, and (2) the number of SIMs of the telecommunication operator in mobile phones connected to a foreign network as a proxy for outgoing international mobility. Other devices (e.g., modems, tablets, wearable devices, etc.) are excluded from this study. In this way, we can quantify both incoming and outgoing international mobility almost in real-time (e.g., with one day of delay).

\subsection{Modeling International Mobility}
\label{sec:modeling}

In this Section, we highlight how we can model international mobility patterns with roaming traces. 
In the literature, there are mainly two ways to model mobility flows: the gravity model \cite{zipf1946p}, and the radiation model \cite{simini2012universal}. The main differences are that the gravity model mimics Newton's gravity law and assumes that the number of trips decreases as the distance between places increases. In this model, the population of the origin and the one of the destination play the role of masses.
The radiation model \cite{simini2012universal}, similarly to the intervening opportunities model \cite{stouffer1940intervening}, assumes that the number of trips is justified by the opportunities offered by the origin and destination locations with people that will eventually travel to a location that can provide adequate opportunities within a certain distance. 

\subsubsection{Gravity Model}
\label{sec:gravity}
In 1946 George K. Zipf proposed a model to estimate mobility flows, drawing an analogy with Newton's law of universal gravitation \cite{zipf1946p}. 
The gravity model is based on the assumption that the number of travelers between two locations increases with the population living there while decreases with the distance between them \cite{barbosa2018human}.   
Given its ability to generate spatial flows and traffic demand between locations, the gravity model has been used in various contexts such as transport planning \cite{erlander1990gravity}, spatial economics \cite{prieto2018scaling}, and the modeling of epidemic spreading patterns \cite{balcan2009multiscale, dudas2017virus, kraemer2019utilizing}. 
In particular, the gravity model estimate mobility flows between the areas $i, j$ according to the following function
\begin{equation}
    T_{i,j} \propto m_i, m_j f(r_ij)
\end{equation}
where the masses $m_i$ and $m_j$ are related to people in location $i$ and $j$ respectively, while $f(r_{ij})$ is a function of the distance between $i,j$ and it is commonly called friction factor or deterrence function. 
There are two common ways to model the deterrence function, namely (i) assuming an exponential decay:
\begin{equation}
    f(r_{ij}) = exp^{- \beta r_ij}
\end{equation}
or (ii) assuming a power decay of the flows with respect to the distance:
\begin{equation}
    f(r_{ij}) = r_{ij} ^ {-\beta}
\end{equation}
The parameters of the function need to be fine-tuned. In this work, we have searched the best parameters using the curve fit utilities of SciPy \cite{virtanen2020scipy}.
The main limitations of the gravity models are (i) that it requires, at least, the estimation and calibration of beta, which makes it sensitive to its changes; and (ii) that for doing this calibration, the system needs empirical data of the actual movements which are not necessarily
available for all cases. As a result of the previous limitations, this approach is a strong simplification of the actual flows, so the results may not reflect the real mobility.

\subsubsection{Radiation Model}
\label{sec:radiation}
To solve some of the limitations of the gravity model, the radiation model has been proposed \cite{simini2012universal}.
This model is an extension of the intervening opportunities model \cite{stouffer1940intervening} in which we assume that a traveler chose the destination of a trip by computing two actions. First, all the possible destinations are assigned to a value representing the opportunities for the traveler. This number $k$ is chosen from a distribution $p(k)$ representing the quality of the opportunity. Then, all the opportunities are ranked according to the distance and the traveler goes to the nearest location with an opportunity value higher than a threshold. The threshold is randomly sampled by the same distribution $p(k)$. Therefore, the number of people commuting from $i$ to $j$ can be modeled with 
\begin{equation}
    T_{ij} = \frac{m_i m_j}{(m_j + s_{ij})(m_i + m_j + s_{ij})}
\end{equation}
and, differently from the gravity model, there are no parameters to calibrate. The radiation model has been reported to better captures long-term
migration patterns and to have an high degree of accuracy at the intra-country scale \cite{simini2012universal, isaacman2018modeling}. The radiation model we adopted is implemented in scikit-mobility library \cite{pappalardo2019scikit}.

Although the radiation model has been applied efficiently in various settings, some results highlight that the spatial scale is not adequately considered by the model \cite{kang2015generalized, masucci2013gravity}. In that sense, some studies go further and limit the application of the radiation model to urban or metropolitan areas \cite{yan2014universal}, due to the parameter-free design of the model, which limits the capability of capturing human mobility. 

\subsection{COVID Gravity Model}
\label{sec:adaptation}
In this work, we claim that the gravity model may have some limitations when modeling human mobility during the COVID-19 pandemic. 
In particular, the gravity model assumes that flows of people are proportional to the population and the distance between origins and destinations. However, during the COVID-19 pandemic we should also consider that travel restrictions and travel bans play an important role. Indeed, if we suppose to have an origin and two different destinations with the same population and the same distance, by definition, the gravity model will output the same flow of people. However, the destinations may have different restrictions in place (e.g., quarantines, travel bans) and thus the flows may be significantly different. Therefore, we claim that capturing only distances and populations is not enough and that the restrictions should be explicitly taken into consideration.

In this Section, we adapted the gravity model to take into consideration also restriction levels. This version of the gravity model is called COVID Gravity Model (CGM).  

The information about restriction levels are provided by the Oxford Stringency Index (SI) \cite{hale2020variation}. It is a composite measure based on nine response indicators including school closures, workplace closures, and travel bans. Oxford SI is provided with different spatial aggregations including the national one and it take values from 0 to 100 where lower numbers indicate lower restrictions. Oxford SI is computed every day starting from the 22$^\text{nd}$ January 2020. As this study is focused on European countries, we investigate a period that goes from the 5$^{th}$ of March to the $30^{th}$ of May. Indeed, starting from March 5, European countries start to adopt non-pharmaceutical interventions to contrast the diffusion of the pandemic (e.g., school closure in Italy and self-isolation in Germany)

CGM considers, additionally to populations and distances, the Oxford SI of the origin country and the Oxford SI of the destination. 

Mathematically, we can model $T_{i,j}$ of CGM as a negative binomial regression with multiple parameters to fit \cite{crymble2018modelling}:

\begin{equation}
    T_{i,j} = exp(\epsilon + \alpha log(P_i) + \beta log(P_j) + \gamma log(f(r_{ij})) + \delta_1 SI_i + \delta_2 SI_j)
\end{equation}

\subsection{Evaluation Metrics}
\label{sec:evaluation_metric}
The S{\o}rensen-Dice index, also called Common Part of Commuters (CPC) \cite{barbosa2018human,luca2021survey}, is a well-established measure to compute the similarity between real flows, $y^r$, and generated flows, $y^g$: 
\begin{equation}
    CPC = \frac{2 \sum_{i,j} min (y^g(l_i, l_j), y^r(l_i, l_j))}
    {\sum_{i,j} y^g(l_i, l_j) + \sum_{i,j} y^r(l_i, l_j)} 
\end{equation}
CPC is a positive number and contained in the closed interval $(0, 1)$ with 1 indicating a perfect match between the generated flows and the ground truth and 0 highlighting bad performance. Note that when the generated total outflow is equal to the real total outflow CPC is equivalent to the accuracy, i.e., the fraction of trips' destinations correctly predicted by the model.
In this work, we use CPC to evaluate the goodness of gravity, radiation and CGM

{ We also compute the Information Gain (IG). Given the real flow at a given time step over $n$ locations $y^r = \{y_1^r,y_2^r,\dots,y_n^r\}$ and the generated flows for the same spatial and temporal reference $y^g = \{y_1^g,y_2^g,\dots,y_n^g\}$, IG is defined as follows

\begin{equation}
    IG(y^r,y^g) = \sum_{i=1}^n \frac{y_i^r}{N} log \frac{y_i^r}{y_i^g}
\end{equation}
where $N$ is the sum over all the elements in $y^r$. IG is a non-negative error metric with lower numbers indicating better performances. We use the Information Gain implemented in scikit-mobility \cite{pappalardo2019scikit}.
}
\section{Results}
\label{sec:results}
In this Section, we first assess the synchrony of mobile phone data and air traffic statistics to validate the collected data (Section \ref{sec:assessing}). Afterwards, we discuss the results obtained in terms of CPC using the gravity and the radiation models (see Section \ref{sec:results_evaluation_gr}), and the ones obtained using the COVID Gravity Model (see Section \ref{sec:cgm_results})

\subsection{Assessing Synchrony with Air Traffic}
\label{sec:assessing}

Here, we show that roaming data generated by mobile phone networks is a good proxy for capturing and modeling international mobility. To this end, we measure the synchrony between the data of international air passengers and the one generated by mobile phone activities.
For the scope of this study, we assume that the number of passengers from air traffic data is representative of incoming/outgoing international mobility in UK. UK's Home Office has recently opened a dataset containing statistics of air passengers' arrivals since the COVID-19 outbreak\footnote{\url{http://bit.ly/AirTrafficStats}}. 

\begin{figure*}[!htb]
	\centering
    \includegraphics[width=1\textwidth]{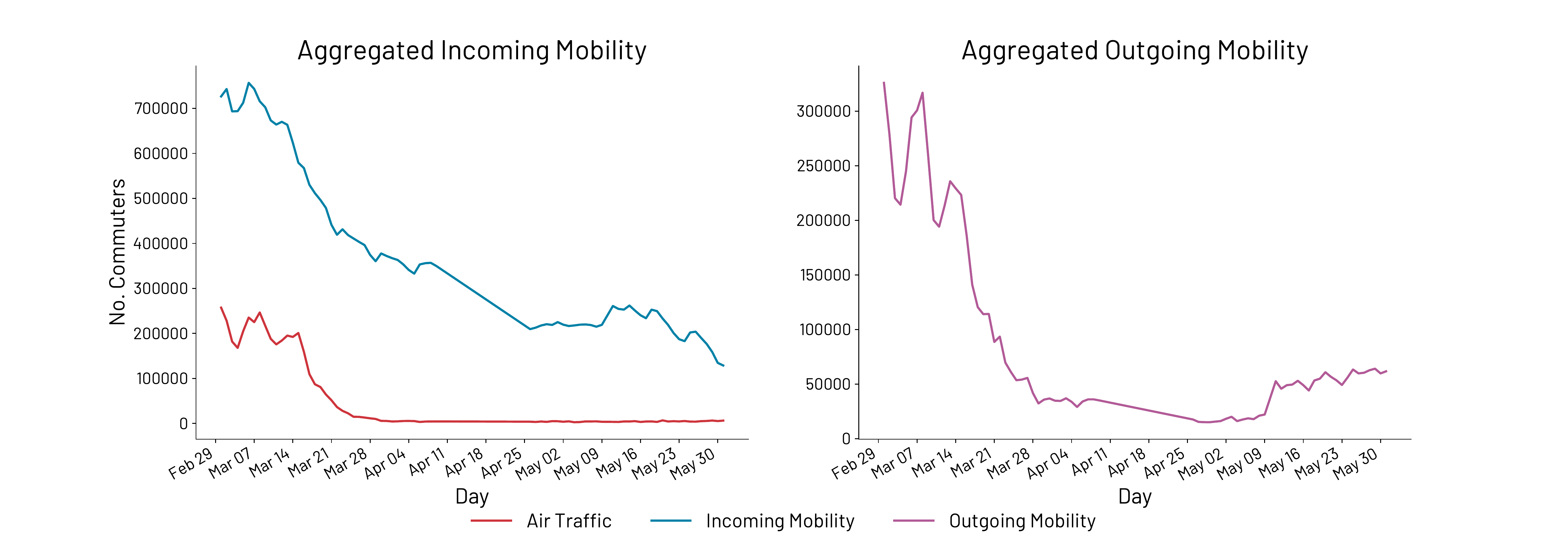}
\caption{\small{On the left, commuters traveling to UK measured with roaming data (blue) and air passengers from air traffic data (red). On the right, commuters from UK to other countries. In both cases, we can spot the effects of the COVID-19 pandemic (e.g., suggestions against all but essential travels, travel bans, lockdown and other countermeasures that impacted international mobility).}}
	\label{fig:in_timeseries}
\end{figure*}

In particular, the dataset details the daily number of air passengers who arrived in UK from January $1^{st}$ 2020 to July $31^{st}$ 2020, obtained from the Advanced Passenger Information (API). The API data primarily relates to passengers coming to UK via the commercial aviation route.
The data is aggregated by day and without considering the origin of the flows. For this reason, to compare the synchronicity of the time series, we aggregate the roaming data without considering the origin. In particular, for each country $c \in C$, we indicate its relative flow to UK at time $t$ as $c_t$. Then, the aggregated flow at time $t$ is 
$$ a_t = \sum_{c \in C} c_t $$
Figure \ref{fig:in_timeseries}, on the right, presents the temporal series of daily air passengers' arrivals (in black) and the ones of aggregated roaming data (in blue) regarding the incoming mobility (i.e., people traveling to UK). On the left, we have the data generated by roaming activity for outgoing international mobility (i.e., people traveling from UK).  

Before assessing the synchronicity, we make a few observations around Figure \ref{fig:in_timeseries}.
First of all, air passengers measure the daily arrivals while mobile phone data measures the presence of international devices (i.e., with a SIM card registered outside UK) roaming on the network. 
Second, according to the data of the Border and Immigration Transactions, the majority of international travelers arriving in UK before April 2020 were traveling by air, while for April and May 2020, the air passengers accounted for only  46\% and 38\% of international travelers going to UK\footnote{\url{http://bit.ly/UK-Stats-Mob}}. 
Finally, as expected, by looking at the trends, we can see how international mobility was deeply affected, both in terms of incoming and outgoing international mobility,  by the limitations imposed as a result of the spread of  COVID-19. For instance, we can see how the three plots start to decrease as mobility restrictions were introduced.
This information provides two important insights on (i) why the number of roaming devices is higher than the one of air passengers and (ii) why starting from the lockdown announced on the $23^{rd}$ of March the two lines related to incoming mobility started to decrease differently.

{ There are many ways to assess the synchronicity of time series with peculiarities and limitations. Among them, Pearson correlation can be used to measure how much two time series co-vary over time. Pearson correlation is a measure that expresses linear relationships between variables. It varies between -1 and 1 where the two extremes are perfect correlations (negative and positive respectively) while 0 indicates no correlation.}
There are two types of synchronicity we want to measure and assess: \emph{local synchronicity} ($\rho_l$) and \emph{global synchronicity} ($\rho_g$). The former allows us to understand whether or not the two series evolve in the same way considering a sliding window of $n$ days. The latter provides insights on the behavioral similarity of the temporal series over the entire period considered.
  
Thus, in order to compute the global and local synchronicity of the timeseries we used Pearson correlation. In the first case, we computed the correlation over the entire temporal series while in the second case we used different temporal windows $W$. In particular, given the series of international commuters' volumes provided by air traffic $X_{\text{air}}$ and by mobile phones $X_{\text{mob}}$, we computed the global synchronicity as
\begin{equation} 
\rho_g = \frac{E[(X_{\text{air}} - \mu x_{\text{air}})(X_{\text{mob}} - \mu x_{\text{mob}})]}{\sigma_{X_{\text{air}}} \sigma_{X_{\text{mob}}}} 
\end{equation}

Similarly, $\rho_l$ is computed by applying a sliding window of size $n$ to the two timeseries. In particular, 

\begin{equation} 
\rho_l = \frac{E[(X_{\text{air}}^{t,t+n} - \mu x_{\text{air}}^{t, t+n})(X_{\text{mob}}^{t, t+n} - \mu x_{\text{mob}}^{t, t+n})]}{\sigma_{X_{\text{air}}^{t, t+n}} \sigma_{X_{\text{mob}}^{t, t+n}}} 
\end{equation}

where $X^{t, t+n}$ is the timeseries in the temporal interval $(t, t+n)$.

The results of the experiments are listed in Table \ref{tab:results_sync}.
{ \small
\begin{table}[]
\begin{tabular}{l|lll}
                       & \multicolumn{1}{l}{ $\mu$} & \textbf{median} & \textbf{W} \\ \hline
\textbf{$\rho_g$}        & \multicolumn{2}{l}{0.926}                                                                                 & -          \\ \hline
\multirow{5}{*}{\textbf{$\rho_l$}} & 0.169                                                                & 0.331                               & 5          \\
                       & 0.246                                                                & 0.328                               & 10         \\
                       & 0.377                                                                & 0.379                               & 15         \\
                       & 0.424                                                                & 0.467                               & 20         \\
                       & 0.460                                                                & 0.576                               & 25        
\end{tabular}
\caption{\small{Results of the global and local synchronicity of the two temporal series. The global synchronicity of the two series is extremely high, while the local synchronicity increases as we enlarge the temporal window.}}
\label{tab:results_sync}
\end{table}
}
As we can see, the global synchronicity of the two series is $0.926$, and it indicates almost a perfect synchronicity. Regarding the local synchronicity, the quality depends on the size of the temporal window $W$. Indeed, as we increase the temporal window, the synchronicity between the two time series increases too. For instance, with $W=10$ the median of $\rho_l$ is $0.328$ that increases to $0.576$ with $W=25$. 

The validations carried out are only related to the incoming international mobility, i.e., people traveling to the UK. Roaming data can provide timely and precise insights also on people traveling from the UK to other countries. In Figure \ref{fig:in_timeseries}, on the left, it is possible to see the temporal series related to outgoing international mobility between March $5^{th}$ and May $31^{st}$ 2020. Unfortunately, we have not validated the temporal series with other statistics as the ones we found for outgoing international mobility involving the UK were monthly aggregated, leading to a temporal series of only three elements.

\subsection{Gravity and Radiation Models' Performances}
\label{sec:results_evaluation_gr}
In this Section, we evaluate the performances of the gravity model and the radiation model both for the incoming and outgoing international mobility flows.

\begin{figure}[!htb]
	\centering
    \includegraphics[width=0.9\columnwidth]{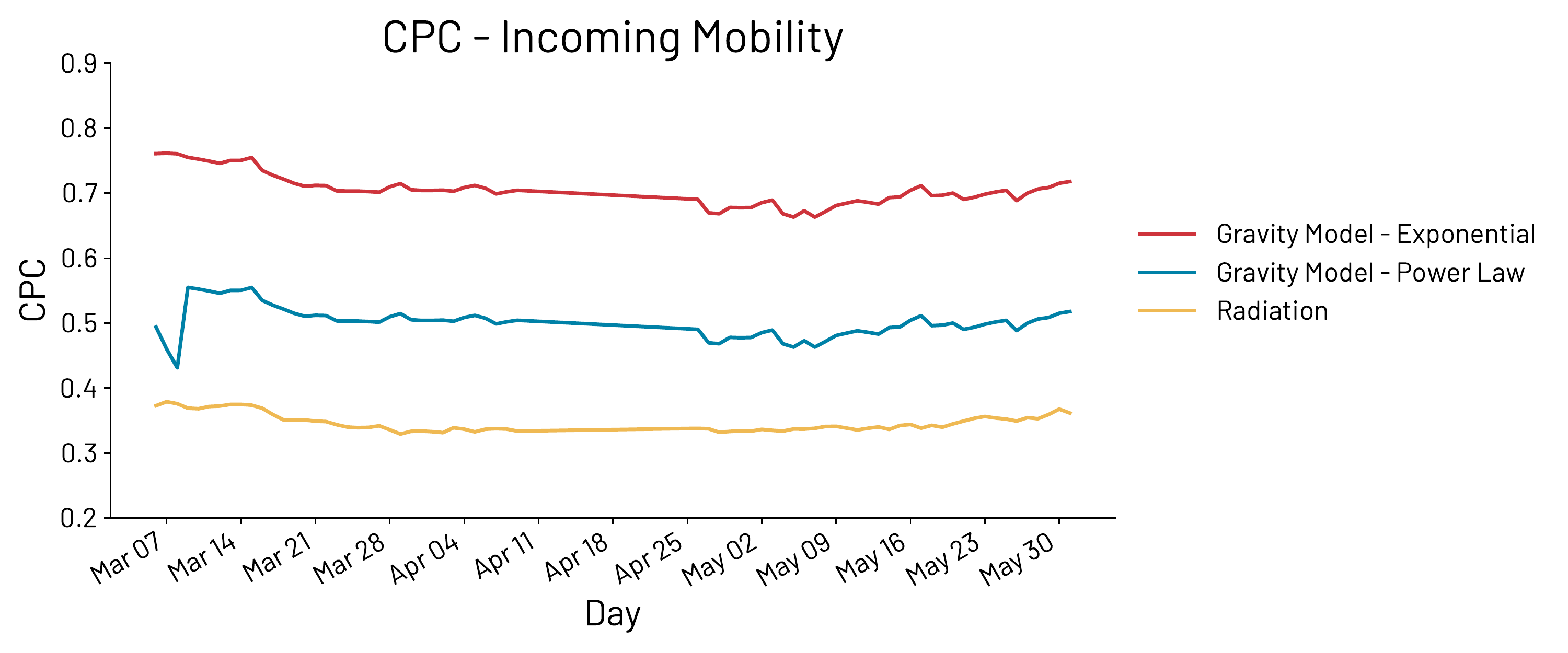}
\caption{\small{CPC related to incoming mobility of the radiation model (in yellow), of the gravity model with an exponential decay (in red), and of the gravity model with a power-law decay (in blue). The model performing better is the gravity model with exponential decay, and this is valid both before and during the introduction of non-pharmaceutical interventions due to the COVID-19 pandemic. In particular, the gravity model with exponential decay reaches an average CPC of 0.762 before the introduction of non-pharmaceutical interventions, 0.666 during the introduction of non-pharmaceutical interventions, and 0.685 over the entire period under analysis.}}
	\label{fig:inflow_performances}
\end{figure}

In Figure \ref{fig:inflow_performances}, we can see how the gravity model with exponential decay achieves the best performances with respect to the other models. A summary of the results is shown in Table \ref{tab:summary_in}. The average CPC of the gravity model with exponential (G-Exp) decay is 0.685, while with a power law (G-Pow) decay, the same model achieves a CPC of 0.448. The worst performing model is the radiation model (R) that has an average CPC of 0.348. These results are in line with the previously highlighted limitations of the radiation model and its problems in capturing mobility beyond urban scale levels, at least when considering incoming mobility \cite{simini2012universal}. 

An interesting investigation regards the goodness of such models in modeling international mobility before and during the pandemic. 
In this sense, we also compute the average CPC for two different periods. The first one is related to the first { ten} days under analysis: March $5^{th}$ to March $15^{th}$. This period is reported as P1 in Table \ref{tab:summary_in}. The second period (P2) started when international flows decreased as COVID-19 pandemic rapidly spreads across Europe. In particular, this period goes from March $16^{th}$ to the end of June. Also, in these periods, the model outperforming the others is G-Exp. However, while the performances of G-Pow and R remain stable across the periods (e.g., the average CPC decrease of 2\% and 3\% respectively), the average CPC of G-Exp decreased by about 10\% in the second period. 

{In Table \ref{tab:summary_in}, we also report the IG of the models. IG is a non-negative number that can be interpreted as an error with values close to 0 that indicate better performances. As we can see, G-Exp has the lower IG followed by G-Pow. R reaches the worst performances also in terms of IG.}

{ \small
\begin{table}
\begin{tabular}{l|llllll}
\textbf{}                    & $\mu$ CPC & max CPC  & min CPC  & $\mu$ CPC - P1 & $\mu$ CPC - P2 & $\mu$ IG  \\ \hline
\textbf{G-Exp.} &   0.705     &   0.761      & 0.663         &    0.753                 &      0.697  & 6.183           \\
\textbf{G-Pow.}   &    0.501    &   0.554      &     0.431    &       0.523              &      0.497  & 9.254         \\
\textbf{R}             &   0.348     &    0.379     &    0.329     &         0.372            &      0.342  & 17.815        
\end{tabular}
\caption{\small{The results in terms of CPC of the gravity model with exponential decay (G-Exp.), the gravity model with power law decay (G-Pow.), and the radiation model (R). We report the average CPC over the entire period under analysis ($\mu$), the maximum and minimum CPC reached, and the average CPC of the first period (i.e., before the introduction of non-pharmaceutical interventions due to COVID-19 pandemic) and the second period (i.e., during the introduction of non-pharmaceutical interventions due to COVID-19 pandemic). The model that better captures incoming international mobility is G-Exp (0.685) followed by G-Pow (0.448) and R (0.348).{ We also report the average IG for each model ($\mu$ IG). Number closer to 0 indicate better performances. Thus, the best performances are achieved by G-Exp. (6.183), followed by G-Pow (9.254) and, finally, the worst performances are reached by R (17.815).} }}
\label{tab:summary_in}
\end{table}
}

\begin{figure}[!htb]
	\centering
    \includegraphics[width=0.9\columnwidth]{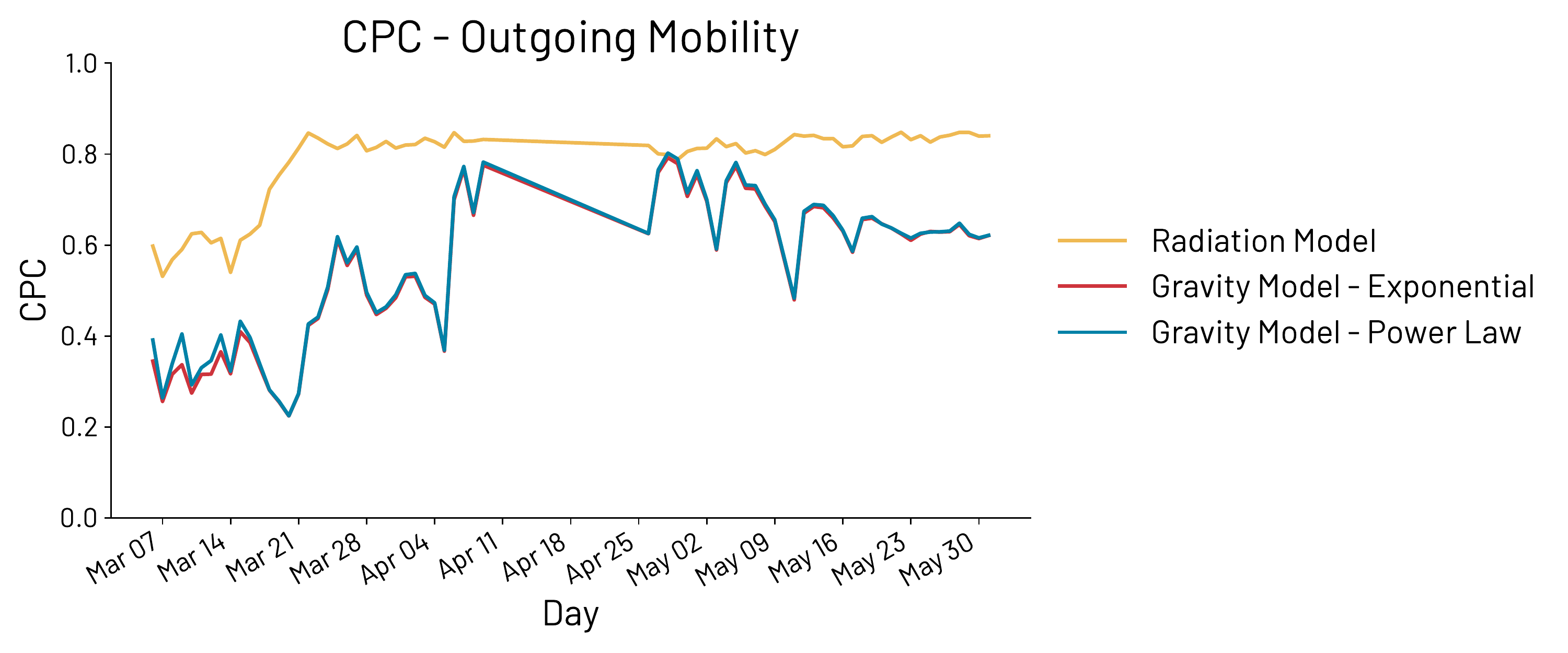}
\caption{\small{CPC of the three models on the outgoing mobility. Radiation model (yellow), gravity with exponential decay (red) and gravity with power law decay (blue) have similar performances in the first period while the best model during the introduction of non-pharmaceutical interventions is the radiation model (average CPC 0.815).}}
	\label{fig:outflow_performances}
\end{figure}

{\small
\begin{table}
\begin{tabular}{l|llllll}
\textbf{}                    & $\mu$ CPC  & max CPC  & min CPC & $\mu$ CPC - P1 & $\mu$ CPC - P2 & $\mu$ IG \\ \hline
\textbf{G-Exp.} &   0.551     &   0.792      & 0.224         &    0.325                 &      0.588     &   12.549  \\
\textbf{G-Pow.}   &    0.557    &   0.802      &     0.224    &       0.352              &      0.591     &  11.386    \\
\textbf{R}             &   0.783     &    0.848     &    0.531     &         0.591            &      0.815   &  4.732      
\end{tabular}
\caption{\small{CPC of the three models when dealing with outgoing mobility. Differently from the incoming mobility, the outgoing mobility before the introduction of non-pharmaceutical interventions can be modeled with all the three approaches with similar performances. However, in the second period, the performances of the radiation model raise up to a CPC of 0.815 while both the gravity models' performances decrease to 0.183 and 0.254. {Concerning IG, the best performances are reached by R (4.732) followed by G-Pow (11.386) and G-Exp (12.549).}}}
\label{tab:summary_out}
\end{table}
}
We carry out the experiments also for outgoing international mobility as shown in Figure \ref{fig:outflow_performances} and Table \ref{tab:summary_out}. The results are extremely different from the ones obtained for the incoming international mobility. In particular, while for the first period P1 going from  March $5^{th}$ to March $15^{th}$ the three models are reliable and provide similar performances, in the second period, when the outgoing mobility dramatically decreased as emerged from Figure \ref{fig:in_timeseries}, the performances of the radiation model improve and reach an 
average CPC of 0.815. On the other hand, the two versions of the gravity models have a drastic drop in the performances moving from an average CPC of 0.547 (G-Exp) and 0.561 (G-Pow) to 0.183 and 0.254, respectively. {In terms of IG, the best performances are achieved by R (4.732) followed by G-Pow (11.386) and G-Exp (12.549).}

The differences between the obtained results when considering incoming and outgoing international mobility are likely influenced by the fact that while the incoming mobility shows a clear and constant trend over the considered period (see Figure \ref{fig:map_flows}), the outgoing mobility is considerably more irregular and thus more challenging to model.

\subsection{COVID Gravity Model Performances}
\label{sec:cgm_results}
\begin{figure}[!htb]
	\centering
    \includegraphics[width=0.9\columnwidth]{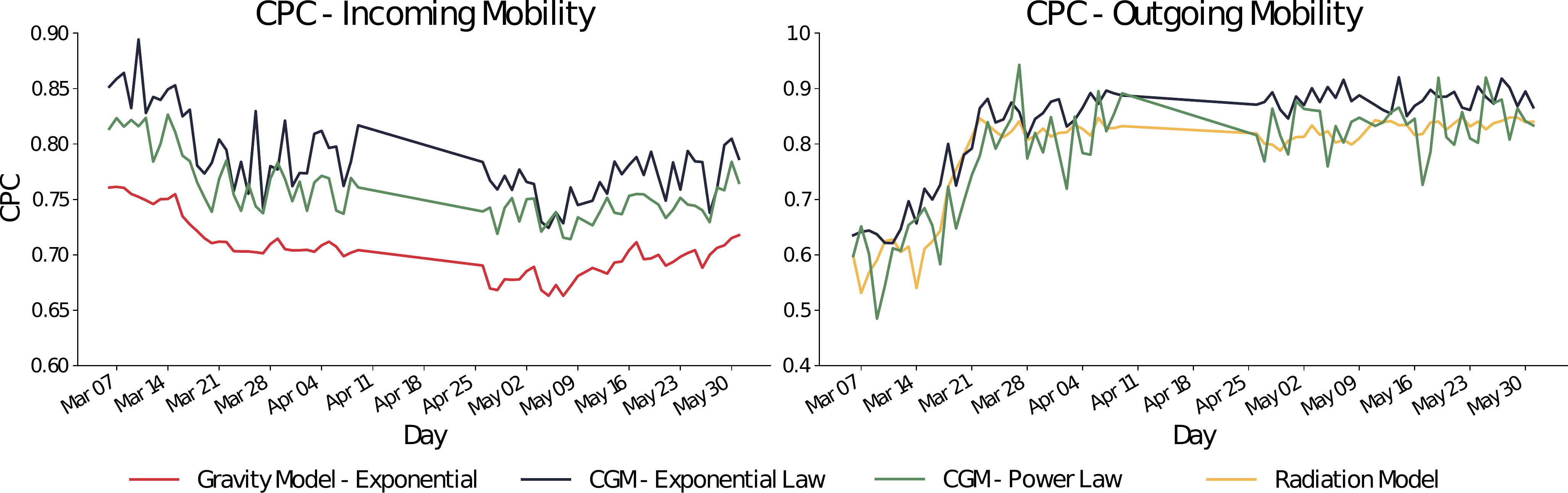}
\caption{\small{On the left, CPC for outgoing mobility of the gravity model with exponential decay (red)  and the two versions of CGM (with exponential law in blue and with power law in green). On the right, the CPC for incoming mobility of the two CGMs and the radiation model (yellow). In both the cases, GCM outperforms the traditional gravity model and, in general, is the model with the higher average CPC.}}
	\label{fig:new_gravity}
\end{figure}

The two versions of the CGM outperform, in terms of CPC, the traditional gravity model independently by the deterrence function. Results are shown in Figure \ref{fig:new_gravity}.

A detailed overview of the average relative improvements is shown in Table \ref{tab:rel_improvement}. Given a value $\hat{y}$ and another value $y$, the relative improvement of $\hat{y}$ over $y$ is computed as $$ rel(\hat{y}, y) = \frac{\hat{y}-y}{y} $$

In this case, we compute the relative improvement for each CPC of CGM over the CPCs of the other models and we report the average relative improvement in Table \ref{tab:rel_improvement}.

\begin{table}[]
\centering

\begin{tabular}{ll|lll}                                        &          & Radiation & Gravity Pow. & Gravity Exp. \\ \hline
\multicolumn{1}{l|}{\multirow{2}{*}{Incoming}} & CGM Exp. & 126.91\%  & 57.13\%      & 11.50\%      \\
\multicolumn{1}{l|}{}                          & CGM Pow. & 118.87\%  & 51.56\%      & 7.55\%       \\ \hline
\multicolumn{1}{l|}{\multirow{2}{*}{Outgoing}} & CGM Exp. & 6.54\%    & 60.89\%      & 63.90\%      \\
\multicolumn{1}{l|}{}                          & CGM Pow. & 0.73\%    & 51.01\%      & 53.76\%     
\end{tabular}%
\caption{The relative improvements of the two versions of CGM with respect to the radiation and gravity models.}
\label{tab:rel_improvement}
\end{table}

In all the scenarios, CGM presents a positive relative improvement with respect to the CPC of the traditional gravity models. Moreover, when modeling the outgoing mobility flows, the model achieving the best performance was the radiation model. By explicitly modeling the mobility restrictions, CGM achieves similar (slightly higher) performances.
More in general, CGM with an exponential decay function is the best way to model both incoming and outgoing international mobility flows during the COVID-19 pandemic. Its average CPC for incoming mobility is 0.78 while for the outgoing mobility is 0.83. In both the cases, CGM with a power law decay function achieves similar performances with a CPC of 0.75 for incoming mobility and 0.78 for outgoing mobility flows. 
With respect to the radiation model, when modeling the incoming flows the performances of CGM are more than double in terms of accuracy (126\% and 118\% more than the radiation model, using an exponential and a power law decay function respectively). On the other hand, when modeling the outgoing mobility CGM performances are similar to the radiation one. In particular, a CGM with a power law decay function outperforms the radiation model by a 0.73\% average relative improvement, while with an exponential function the relative improvement grows up to 6.54\% on average. 
Finally, we have similar relative improvements for the gravity model with a power law decay both in the incoming and outgoing mobility modeling tasks. 

The results obtained can be useful in many scenarios and highlight some important suggestions. First of all, in pandemic times, modeling just the mobility flows is not enough and explicitly modeling the severity of non-pharmaceutical interventions and other policies of the origin and destination countries is fundamental. This is shown by the significant relative improvements obtained with CGM. 
For instance, by using CGM and explicitly modeling restrictions, policy makers can take more precise decisions based on a more accurate model. At the same time, given the strong relation between mobility and disease diffusion, CGM can help in better understanding how a disease circulates internationally. 

\section{Discussion and Limitations}
\label{sec:discussion}
In this Section, we discuss some implications and the limitations of the data source and models used in this study. 

Regarding the data source, we use roaming data generated by mobile phones as a proxy of international mobility. This data source presents some peculiar advantages. In particular, it offers timely insights on mobility flows as data can be easily processed every day. Moreover, the spatial granularity of the data can be significantly fine-grained (e.g., antenna level), and thus policy makers can gather precious insights for taking decisions. For example, having timely and spatially fine-grained data is helpful when we want to analyze the spreading of new COVID-19 variants internationally. Moreover, roaming data may allow to investigate how international travelers move within a foreign country (e.g., antenna level position) and this is an important advantage that only roaming data can offer. 

On the other hand, however, mobile phone data are generally associated with some limitations like the possibility of accessing the data and some other biases e.g., owners of the SIM cards, not possible to correctly monitor people younger than 18 years old and others \cite{luca2021leveraging}.

The usage of roaming data has also some additional limitations. In this study, when we deal with roaming data, we are simply counting how many mobile phone SIM cards registered in another countries are in the UK in a specific day. Therefore, we are also counting people that may live in the UK but, for any reason, have a foreign SIM card. Moreover, when a SIM is roaming data in a foreign country, it is likely to connect to the antennas of multiple different providers. For example, given two telecommunication operators $A$ and $B$, a mobile phone may use services offered by $A$ for a couple of days, then connect to $B$ without leaving the country and finally connect again to $A$'s services. In this study, the mobile phone will be counted as an incoming commuter for two times in two different days even if they never left the country. Data may be also biased by people traveling with more than one SIM card that, in this 
study, are eventually counted multiple times.

Even if roaming data may contain some measurements errors, we validated the temporal correlation of the extracted time series with the ones of international flight statistics obtaining significantly high correlations indicating the potential goodness of the data.

{ As part of future work, it may also be of interest to collect more fine-grained air traffic data and use a data fusion mechanism to leverage the advantages of all the available data sources. At the moment, data fusion mechanisms for human mobility have been used only on city- and regional-scale \cite{toole2015path,noulas2013exploiting,lau2019survey} but may provide some advantages also for capturing international mobility. { At this stage, we did not use any data fusion technique due to the fact that the available air traffic statistics have two significant limitations. First of all, UK's border control provides only the data for people arriving in the UK and there is no information about the people leaving the UK. Second, the data are not aggregated at country level (i.e., we do not know the origin of the travelers). For these two reasons, adopting data fusion techniques in this study would not provide any valuable information as we are working with data with different aggregations, thus representing two (slightly) different phenomena.}}

Concerning the models, in this work we show that the traditional models are not using enough information to model human mobility.
{ Existing solutions are the gravity model and the radiation model and fully rely on distance and population. In this study, we claim that focusing on population and distance is not enough. In particular, in the case of the gravity model, we can not fit the parameters to estimate the impact of populations and distances on the flows. For instance, the flows observed in the training set may be significantly biased according to the restrictions imposed by the origin country and the destination country (e.g., quarantine requirements, international travel restrictions). The estimation of parameters may be significantly affected by such biases and underestimated. The radiation model may partially overcome this limitation as it is parameter-free. However, clearly specify which are the (mobility) restrictions imposed in the various countries may be used to boost the performances. In the proposed solution, we use the Oxford Stringency Index as a proxy of the pandemic situation in origin and destination countries and also as a proxy of international travel restrictions. We have seen that by explicitly modeling the restrictions, the performances of the so-called COVID Gravity Model increase both for incoming and outgoing international mobility flows. Having a more realistic model of international travelers is fundamental to provide decision makers with more realistic simulations and estimations. Policy makers may use such insights for taking actions to contrast the diffusion of the disease or to implement policies for improving the well-being of international migrants. Also, the combination of CGM and the fine-grained spatial aggregations, we may have with mobile phone data, may also be used to better investigate many problems including the spreading of new variants of COVID-19, preferences and habits of international travelers and others. For instance, if we want to study how COVID-19 spread internationally (e.g., \cite{lemey2021untangling,ruktanonchai2020assessing}), the availability of fine-grained mobile phone data may lead to more realistic estimations of international mobility flows, and thus more realistic simulations. Also, while the data sources adopted in most of the studies only allow to estimate the number of people traveling from a country to another, with mobile phone data it is also possible to investigate how international travelers move within a country with a variety of fine-grained spatial resolutions (e.g., antenna level). This may help in better understanding how a disease is diffused over a territory and to target more specific geographical areas with countermeasures.}

We acknowledge there are other more sophisticated models based on deep learning techniques as explained in a recent survey \cite{luca2021survey}. Examples of works that model mobility using deep learning are Deep Gravity \cite{simini2021deep}, SI-GCN (Spatial Interaction Graph Convolutional Network) \cite{yao2020spatial} and GMEL (Geocontextual Multitask Embedding Learner) \cite{liu2020learning}. However, given the quantity of data needed to accurately train these models, we decided not to use them in this study where we have only an egocentric network for UK movements. 

\section{Conclusions}
\label{sec:conclusions}
While human mobility is an active research area both at national and local scales, there are fewer studies regarding international mobility patterns and their challenges (e.g., migration and disease diffusion). Tackling such challenges requires timely data with a proper spatial aggregation that roaming activities can provide.
In this paper, we have proposed to use roaming network data to capture international mobility. 
Then, we use the gravity and radiation models to model international mobility. The incoming mobility is modeled better by a gravity model with an exponential decay both before and during the non-pharmaceutical interventions introduced for contrasting the spread of the COVID-19 pandemic. Instead, the outgoing mobility is captured equally well by the various models before the mobility restrictions were introduced. On the other side, after the second week of March, the radiation model is the one that captures mobility better.
However, by explicitly modeling the COVID-19 restrictions for the origin and destination countries, the COVID Gravity Model (CGM) outperforms all the other models both for the incoming (improvement up to 126.9\%) and outgoing (improvement up to 63.9\%) mobility scenarios. These findings may have significant impact on how we should model international mobility in times of crises and can help policy makers in taking more accurate decisions. As part of future works, we will evaluate CGM also at a national and sub-national scales. 

\begin{backmatter}
\section*{Availability of data and material}
Data and material are not publicly available. The owner of the data is a private telecommunications operator and data cannot be shared. 

\section*{Competing interests}
  The authors declare that they have no competing interests.

\section*{Funding}
  The work of Andra Lutu was supported by the EC H2020 Marie Curie Individual Fellowship 841315 (DICE). 
  
\section*{Author's contributions}
    M.L. designed the model and performed the experiments. E.M. and A.L. directed the study. All authors contributed to interpreting the results and writing the paper. 
  
\section*{Acknowledgements}
    Not applicable
    
\section*{Ethical considerations}
    The data collection and retention at network middle-boxes and elements are in accordance with the terms and conditions of the MNO and the local regulations. All datasets used in this work are covered by NDAs, prohibiting any re-sharing with 3rd parties even for research purposes. Further, raw data has been reviewed and validated by the operator with respect to GPDR compliance (e.g., no identifier can be associated to person), and data processing only extracts aggregated user information at postcode level.
    No personal and/or contract information was available for this study and none of the authors of this paper participated in the extraction and/or encryption of the raw data.

\bibliographystyle{bmc-mathphys} 
\bibliography{bmc_article}      

\end{backmatter}
\end{document}